\documentclass{aa}  

\usepackage{graphicx}
\usepackage{bm}
\graphicspath{{figs/}}

\usepackage{txfonts}
\usepackage[colorlinks,allcolors=blue]{hyperref}
\usepackage{amstext}

\begin{document}

   \title{Origin of eclipsing time variations in post-common-envelope 
   binaries: Role of the centrifugal force}

   \author{Felipe H. Navarrete\inst{1,2}\thanks{felipe.navarrete@hs.uni-hamburg.de}
          \and
          Dominik R.G. Schleicher\inst{3}
          \and
          Petri J. K\"apyl\"a\inst{4,2}
          \and
          Carolina A. Ortiz-Rodr\'iguez\inst{3}
          \and\\
          Robi Banerjee\inst{1}
          \fnmsep
          }

   \institute{Hamburger Sternwarte, Universit\"at Hamburg, Gojenbergsweg 112,
           21029 Hamburg, Germany
            \and
            Nordita,
            Stockholm University and KTH Royal Institute of Technology,
            Hannes Alfvéns väg 12, SE-106 91 Stockholm, Sweden
         \and
            Departamento de Astronom\'ia, Facultad Ciencias F\'isicas y
            Matem\'aticas, Universidad de Concepci\'on, Av. Esteban Iturra
            s/n Barrio Universitario, Casilla 160-C, Concepci\'on, Chile
            \and
            Institut f\"ur Astrophysik und Geophysik, Georg-August-Universit\"at G\"ottingen,
            Friedrich-Hund-Platz 1, 37077 G\"ottingen, Germany
            }

   \date{Received --; accepted --}


  \abstract
  {Eclipsing time variations (ETVs) in post-common-envelope binaries (PCEBs) were proposed 
  to be due to the time-varying component of the stellar gravitational
  quadrupole moment. This 
  is suggested to be produced by changes in the stellar structure due to an 
  internal redistribution of angular momentum and the effect of the 
  centrifugal force. We examined this hypothesis and present 3D 
  simulations of compressible magnetohydrodynamics (MHD) performed with the 
  {\sc Pencil Code}. We modeled the stellar dynamo for a solar-mass star with angular 
  velocities of 20 and 30 times solar. We included and varied the strength of the 
  centrifugal force and compared the results with reference simulations without the 
  centrifugal force and with a simulation in which its effect is enhanced. 
  The centrifugal force causes perturbations in the evolution of the numerical model,
  so that the outcome in the details becomes different as a result of
  nonlinear evolution. While the average density profile is unaffected by the 
  centrifugal force, a relative change in the density difference between high 
  latitudes and the equator of $\sim10^{-4}$ is found. The power spectrum 
  of the convective velocity is found to be more sensitive to the angular 
  velocity than to the strength of the centrifugal force. The quadrupole 
  moment of the stars includes a fluctuating and a time-independent component, 
  which vary with the rotation rate. As very similar behavior is produced in absence of the centrifugal force, we conclude that it is not the 
  main ingredient for producing the time-averaged and fluctuating
  quadrupole moment of the star.
  In a real physical system, we thus expect 
  contributions from both components, that is, from the 
  time-dependent gravitational force from the variation in the quadrupole term 
  and from the spin-orbit coupling that is due to the persistent part of the quadrupole.}

   \maketitle
\section{Introduction}
Eclipsing time variations (ETVs) have been observed in a wide range of
post-common-envelope binaries (PCEBs) \citep{zorotovic13, bours16}.
Traditionally, two explanations have been proposed for the observed variations:
One explanation refers to the possible presence of a third body, preferentially with 
a mass of a few Jupiter masses in the case of NN~Ser \citep{beuermann10}
and a brown dwarf of 0.035$M_\odot$ in V471~Tau \citep{Vaccaro15},
and on a wide orbit, which could explain the observed ETVs 
as due to the orbit of the binary system around the common center of
mass via the light travel time effect \citep[e.g.,][]{beuermann12, beuermann13a}.
The presence of such planets might be explained either because they
survived the common-envelope event \citep{volschow14} or because they formed
from the ejecta of common-envelope material \citep{schleicher14}. However, the 
planetary systems were sometimes found to be unstable \citep{Mustill13}, and at other times,
the predicted planets were not detected \citep{Hardy15}.

An alternative possibility is that the ETVs is caused in 
the binary system itself, as a result of magnetic activity. This might occur in different forms. An early suggestion by \citet{decampli79} 
considered a rocket effect produced by anisotropic mass loss, but the hypothesis was finally rejected. Tidal torques are another possibility, but their 
magnitudes are so low \citep[]{zahn89,OgilvieLin07} that they cannot transfer the necessary 
angular momentum \citep{applegate87}. As a different solution, both
\citet{matese83} and \citet{applegate87} proposed that the orbital period would be 
changed if the stellar quadrupole moment changed as a result of magnetic activity. 

This was a central step, but the 
cause for the change in the stellar quadrupole moment remains to be defined and its strength needs to be determined. 
In the original models \citep[e.g.,][]{matese83, applegate87}, it was assumed 
that the magnetic field deforms the star by causing a deviation from its 
hydrostatic equilibrium, requiring thus a very strong magnetic field. \citet{
marsh90}  showed, however, that the required periodic deformation was
too strong to be 
sustained by the luminosity of the star. A different scenario thus emerged 
in which the change of the quadrupole moment is not caused directly by the 
magnetic field, but is a result of angular momentum redistribution inside the 
star through the dynamo process, which then leads to stellar distortions as a 
result of the centrifugal force \citep{applegate92}. Within a simplified
thin-shell model, considering an inner core and an infinitely thin shell,
\citet{applegate92} calculated the quadrupole moment of the shell as
\begin{equation}
        Q=\frac{1}{9}M_sR^2\left(\frac{\Omega^2R^3}{GM}  \right),
\end{equation}
where $M$ is the mass of the star, $M_s\ll M$ is the mass of the
shell, $\Omega$ is the angular velocity of the shell, $R$ 
is the radius from the center of the
star to the shell, and $G$ is the gravitational constant.
\citet{applegate92} calculated the angular momentum to be 
transferred within the star to produce a period variation $\Delta P$ as
\begin{equation}
        \Delta J=-\frac{GM^2}{R}\left( \frac{a}{R} \right)^2\frac{\Delta P}{6\pi},
\end{equation} 
with $a$ the separation of the binary system. The energy required to transfer the 
angular momentum is then
\begin{equation}
        \Delta E=\Omega_{dr}\Delta J+\frac{(\Delta J)^2}{2I_{\rm eff}},
\end{equation}
where $\Omega_{dr}$ refers to the difference of angular velocity between the 
shell and the core, and $I_{\rm eff}$ is the effective moment of inertia, 
corresponding to about half of the inertial moment of the shell.

This model was extended by \citet{brinkworth06}, who considered a finite 
instead of an infinitely thin shell, showing that the latter 
increased the energy required to produce the deformation by roughly one order 
of magnitude. \citet{Voelschow16} subsequently applied this model in a 
systematic way to the sample of \citet{zorotovic13}, showing that the energy 
requirement is sometimes fulfilled and sometimes it is not. A similar conclusion was 
obtained by \citet{Navarrete18} through an extension of the analysis. On the 
other hand, more detailed 1D models solving the evolution equation for the 
stellar angular momentum indicated that the energetic requirement may actually 
be reduced \citep{Lanza06}, and \citet{Voelschow18} concluded that the 
mechanism is feasible for stars with masses of $0.3-0.36$~M$_\odot$. 

Other types of solutions have also been proposed. For instance,
\citet{Applegate89} derived librating and circulating solutions in the 
presence of a constant quadrupole moment, although they were originally 
predicted to provide modulations over shorter periods than observed in the 
PCEBs. \citet{Lanza20} re-examined this scenario, however, and proposed that a 
persistent nonaxisymmetric internal magnetic field could produce an 
appropriate quadrupole moment to explain the observed deviations at a much 
lower energetic expense than in the scenario in which the quadrupole moment 
variation is produced via the centrifugal force \citep[e.g.,][]{applegate92}.

This problem was recently revisited in 3D magnetohydrodynamical (MHD) simulations, 
and although the centrifugal force was not included, quasi-periodic 
quadrupole moment variations caused by magnetic activity were found. They were roughly still 
one order of magnitude lower than required by observations, however \citep{Navarrete20}.
As they were not driven via the centrifugal force, it seems more 
likely that a change in the internal circulation in the star rather than a 
redistribution of the angular momentum has caused this result. This was 
recently confirmed via an extended set of simulations with a more detailed 
analysis \citep{Navarrete22}.

As the correct mechanism that gives rise to the ETVs is 
still not established, it is fundamental to investigate how the 
centrifugal force influences the change in the quadrupole moment within the 
stars. For this purpose, we present 3D MHD 
simulations of a solar-mass star that include and vary the centrifugal force
to assess in this way how it affects the stellar structure. Thus, 
we aim to verify whether the origin of these variations is based on the centrifugal 
force as proposed by \citet{applegate92}, or if other mechanisms must be at 
play to cause the observed variations. Our numerical approach is presented in 
section 2, and the results are given in section 3. We finally present our 
discussion and conclusions in section 4.

\section{Model}\label{sec:model}

We present two sets of simulations with rotation rates $20\Omega_\odot$ and $
30\Omega_\odot$, where $\Omega_\odot$ is the solar rotation rate. These are
part of an overall larger set of simulations that 
has been pursued to analyze dynamos in the context of 
young stars (Navarrete et al., in prep.). We label the first set
simulation group C and the second set group D. Within each set, we varied the 
centrifugal force amplitude.

The compressible MHD equations were solved on a spherical
grid with coordinates $(r,\theta,\phi),$ where $0.7 \leqslant r \leqslant R$ is 
the radial coordinate, $R$ is the radius of the star, $\pi/12\leqslant \theta \leqslant
11\pi/12$ is the colatitude, and $0\leqslant \phi < 2\pi$ is the longitude. The
model is the same as in \citet{Kapyla13} and \citet{Navarrete20, Navarrete22}.
The equations were solved in the following form:
\begin{align}
       & \frac{\partial\vec{A}}{\partial t} = \vec{u}\times\vec{B} -
        \eta\mu_0\vec{J}, \\
        & \frac{{\rm D}\ln \rho}{{\rm D}t} = -\bm\nabla\bm\cdot \vec{u},\\
       & \frac{{\rm D}\vec{u}}{{\rm D}t} = \vec{{\cal F}}^{\rm grav} +
        \vec{{\cal F}}^{\rm Cor} + \vec{{\cal F}}^{\rm cent} -
        \frac{1}{\rho}(\bm\nabla p - \vec{J}\times\vec{B} - \bm\nabla\bm\cdot2\nu\rho
        \vec{S}), \\
       & T\frac{{\rm D}s}{{\rm D}t} = \frac{1}{\rho}\left[\eta\mu_0\vec{J}^2 -
          \bm\nabla\bm\cdot(\vec{F}^{\rm rad} + \vec{F}^{\rm SGS})\right] + 
        2\nu\vec{S}^2,
\end{align}
where $\vec{A}$ is the magnetic vector potential, $\vec{B}=\nabla\times\vec{A}$
is the magnetic field, $\vec{u}$ is the velocity field, $\eta$ is the magnetic
diffusivity, $\mu_0$ is the vacuum permittivity, $\vec{J}$ is the current
density, $\rho$ is the density, $p$ is the pressure, $\nu$ is the viscosity,
$\vec{S}$ is the rate of strain tensor, $T$ is the temperature, and $s$ is the
entropy. $\vec{F}^{\rm rad}$ and $\vec{F}^{\rm SGS}$ are the radiative
and the subgrid scale fluxes, respectively
\citep[see, e.g.,][]{Kapyla13}.
The SGS flux is given by
\begin{equation}
\vec{F}^{\rm SGS} = - \chi_{\rm SGS} \rho T \bm\nabla s,
\end{equation}
where $\chi_{\rm SGS} = \chi^{\rm m}_{\rm SGS} = 0.4\nu$ at $0.75<r/R<0.98$ and
increases smoothly to $12.5 \chi^{\rm m}_{\rm SGS}$ above $r=0.98R$. Below
$r=0.75R$, it decreases smoothly and approaches zero. This term
is a parameterization of the unresolved turbulent heat transport. The
SGS diffusivity is needed because the radiative diffusivity
$\chi=K/c_{\rm P} \rho$, where $K$ is the heat conductivity and
$c_{\rm P}$ is the specific heat at constant pressure, is insufficient
to smooth grid-scale fluctuations even with the enhanced luminosity of
the current simulations.
Furthermore,
\begin{align}
       & \vec{{\cal F}}^{\rm grav} = -(GM/r^2)\hat{\vec{r}},\\
       & \vec{{\cal F}}^{\rm Cor}  = -2\vec{\Omega}_0\times\vec{u}, \\
       & \vec{{\cal F}}^{\rm cent} = -c_f\vec{\Omega}_0\times(\vec{\Omega}_0
        \times\vec{r})
\end{align}
are the gravitational, Coriolis, and centrifugal forces. Here, 
$\vec{\Omega}_0$ is the rotation rate of the modeled star. The parameter $c_f$ was
introduced by \citet{Kapyla20b} and controls the strength of the centrifugal
force. $c_f = 1$ corresponds to the unaltered centrifugal force amplitude, and
$c_f = 0$ implies no centrifugal force. It is defined as
\begin{equation}
        c_f = \frac{\left|\vec{{\cal F}}^{\rm cent}\right|}
        {\left|\vec{{\cal F}}^{\rm cent}_0\right|}
,\end{equation}
with $|\vec{{\cal F}}^{\rm cent}_0|$ being the physically consistent
magnitude of the
centrifugal force.  The need to control the centrifugal force arises due to
the enhanced luminosity and rotation rate in simulations of stellar (magneto-)
convection. This approach is necessary to avoid a too large gap
between acoustic, convective, and thermal relaxation timescales
in simulations that solve the compressible MHD equations
\citep[e.g.,][]{Brandenburg2005, Kapyla13}.
Using the realistic stellar luminosity would have the consequence that
flow velocities would be much lower than the sound speed. The time step would 
then become prohibitively short and the thermal relaxation
(Kelvin-Helmholtz) time
prohibitively long \citep[see]
[for the effects of varying luminosity on the flow properties]{Kapyla20b}.
The enhancement of the luminosity is described by
\begin{equation}
        \mathcal{L}_r=\frac{\mathcal{L}_{\rm sim}}{\mathcal{L}_*},
\end{equation}
where ${\cal L}_{\rm sim}$ is the luminosity in our model and
${\cal L}_*$ is the luminosity of the target, the physical star.
We have ${\cal L}_r=8.07\times10^5$ for the setup adopted here for a
solar-like target star
\citep{Navarrete20}. The angular velocity has to be enhanced correspondingly 
to produce a realistic Coriolis number. For 
the centrifugal force, on the other hand, the strength should be limited
so that the impact on the structure 
of the star is not overestimated. For numerical stability and as outlined by \citet{Kapyla20b}, 
each run was initialized with $c_f=0$, and it was increased in small 
incremental steps after the saturated regime was reached. In this way, the 
effect of the centrifugal force can be explored in the simulation.

To quantify the strength of the centrifugal force, we computed the ratio
of gravitational to centrifugal forces in the simulations presented here as 
well as for a real Sun-like star with the same rotation rate. The ratio of the 
two is defined as
\begin{equation}
        \mathcal{F} = \frac
                       {(|\vec{{\cal F}}^{\rm cent}|/
                       |\vec{{\cal F}}^{\rm grav}|)_{\rm sim}}
                       {(|\vec{{\cal F}}^{\rm cent}|/
                       |\vec{{\cal F}}^{\rm grav}|)_\star},
\end{equation}
where the subscript asterisk$\text{}$ denotes the real star and {\it sim} the simulations.
If this ratio is equal to unity, the relative
strength of the centrifugal force with respect to gravity is the same
in the simulation as in the real star. Particularly for
rapidly rotating stars, it is in principle harder to model a case with
$\mathcal{F}=1$, and we typically remain somewhat below this ratio, but we 
also present a case with $\mathcal{F}>1$ for comparison.

\begin{table*}
\caption{Summary of the dimensionless parameters that characterize the simulations.}
\label{tab:runs}
\centering
\begin{tabular}{c c c c c c c c c c c c c c c c}
  \hline\hline
  Run & $\Omega/\Omega_\odot$ & $c_f$ & Co & Ta & Re & Re$_{\rm M}$
      & Pe & $\mathcal{F}$ & $\langle\Delta_\Omega^{(r)}\rangle_t$ &
      $\langle\Delta_\Omega^{(60^\circ)}\rangle_t$
      & $\Delta_{\Omega{\rm, rms}}^{'(r)}$ & $\Delta_{\Omega{\rm, rms}}^{'(60^\circ)}$
        \\
  \hline
  C1 & 20 & 0       & 57.2  & 2.53(9) & 22.1 & 22.1 & 55.3 & 0        &  9.16(-4) & 3.42(-3) & 4.26(-4) & 5.02(-4) \\
  C2 & 20 & 1.0(-4) & 55.6  & 2.53(9) & 22.8 & 22.8 & 57.0 & 8.75(-1) &  9.27(-4) & 3.60(-3) & 4.62(-4) & 5.30(-4) \\
  C3 & 20 & 1.0(-4) & 56.7  & 2.53(9) & 22.3 & 22.3 & 55.8 & 8.75(-1) &  8.40(-4) & 3.66(-3) & 6.23(-4) & 6.28(-4) \\
  D1 & 30 & 0       & 137.9 & 5.72(9) & 13.8 & 13.8 & 34.6 & 0        & -2.51(-4) & 4.51(-4) & 8.05(-5) & 1.23(-4) \\
  D2 & 30 & 1.0(-4) & 137.8 & 5.72(9) & 13.8 & 13.8 & 34.8 & 8.75(-1) & -2.42(-4) & 4.75(-4) & 1.15(-4) & 1.54(-4)\\
  D3 & 30 & 1.0(-3) & 130.1 & 5.72(9) & 14.6 & 14.6 & 36.6 & 8.75     & -6.34(-5) & 8.70(-4) & 1.52(-4) & 2.39(-4) \\
  \hline
  \end{tabular}
  \tablefoot{Co is the Coriolis number, Ta is the Taylor number, Re and Re$_{\rm M}$
  are the fluid and magnetic Reynolds numbers, and Pe is the P\'eclet number.
  For each run, Pr $ = 60$, Pr$_{\rm M} = 1$, and Pr$_{\rm SGS} = 2.5$.
  $\Delta_\Omega^{(r),(60^\circ)}$ denote the radial and latitudinal
  differential rotation and are defined in Eqs.~(\ref{eq:nablar}) and
  (\ref{eq:nablalat}), respectively. $\langle$...$\rangle_t$ denotes averages
  over time, and the prime denotes fluctuating quantities.}
\end{table*}

The details of the model are further described in
\citet{Navarrete20, Navarrete22} and in \citet{Kapyla13}, and we refer to these papers to avoid 
 repetition. We nonetheless recall that the model assumes an outer 
spherical boundary at the stellar radius that is assumed to be impenetrable 
and stress-free. At the lower boundary at $70\%$ of the stellar radius, the 
magnetic field is assumed to obey a perfect conductor boundary condition, while at the 
top boundary, the field is assumed to be radial. The temperature gradient is fixed at 
the bottom, while a blackbody condition is applied at the surface. The setup includes 
colatitudinal boundaries at $15^\circ$ and $165^\circ$, which are
assumed to be stress-free and perfectly conducting. Density and entropy are 
assumed to have zero first derivatives on colatitudinal boundaries. The
gravitational potential is spherically symmetric and independent of time,
and self-gravity is not taken into account.
The equations are solved with the {\sc Pencil Code}\footnote{https://github.com/pencil-code/pencil-code}, a high-order finite-difference code for compressible MHD equations
\citep{PencilCode}.

We define the Coriolis, Taylor, Reynolds, magnetic Reynolds, Prandtl, magnetic
Prandtl, SGS Prandtl, and P\'eclet numbers
as
\begin{gather}
{\rm Co} = \frac{2\Omega_0}{u_{\rm rms} k_1}, \
{\rm Ta} = \!\left[ \frac{2\Omega_0(0.3R)^2}{\nu}\right]^2,\
{\rm Re} = \frac{u_{\rm rms}}{\nu k_1},\\
{\rm Re_M} = \frac{u_{\rm rms}}{\eta k_1}, \
{\rm Pr} = \frac{\nu}{\chi_m},\
{\rm Pr_M} = \frac{\nu}{\eta},\
{\rm Pr_{\rm SGS}} = \frac{\nu}{\chi_{\rm SGS}^{\rm m}},\\
{\rm Pe} = \frac{u_{\rm rms}}{\chi_{\rm SGS}^{\rm m}k_1},
\end{gather}
where $u_{\rm rms}$ is the root-mean-square velocity, $k_1 = 2\pi/0.3R$ is an
estimate of the wavenumber of the largest convective eddies, and
$\chi_{\rm SGS}^{\rm m} = 0.4\nu$ is the subgrid-scale entropy diffusion in the middle
of the convective region. Each run is characterized by ${\rm Pr} = 60$,
${\rm Pr}_{\rm M} = 1$, and ${\rm Pr}_{\rm SGS} = 2.5$. The other quantities are
shown in Table~\ref{tab:runs}. Throughout this paper, overbars denote
averages over longitude.

\section{Results}\label{sec:results}

\begin{figure}
\centering
\includegraphics[width=\columnwidth]{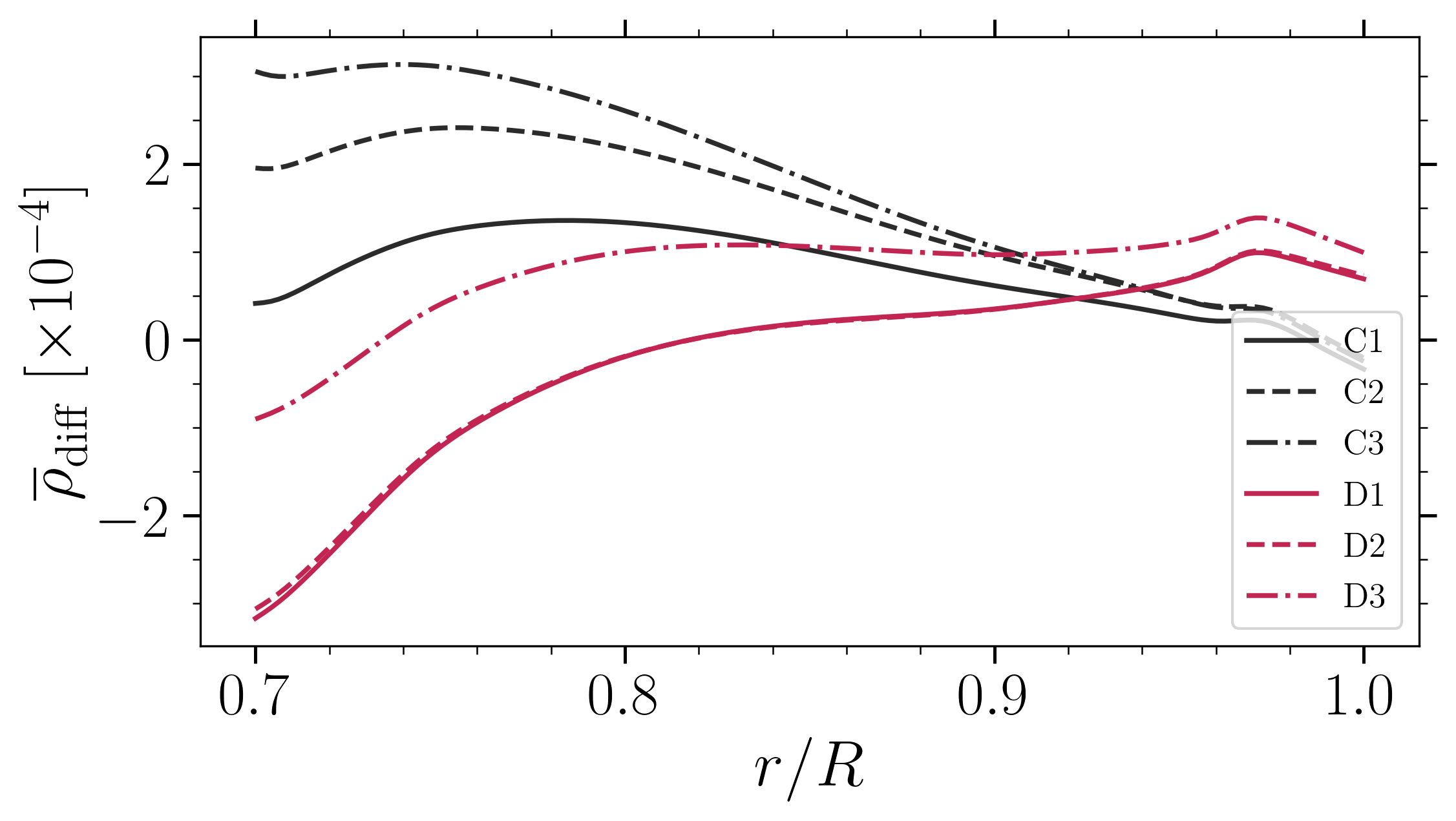}
\caption{Density difference between regions at $60^\circ$ above the equator and
the equator.}
\label{fig:rhodiff}
\end{figure}

\begin{figure}
        \centering
        \includegraphics[width=\linewidth]{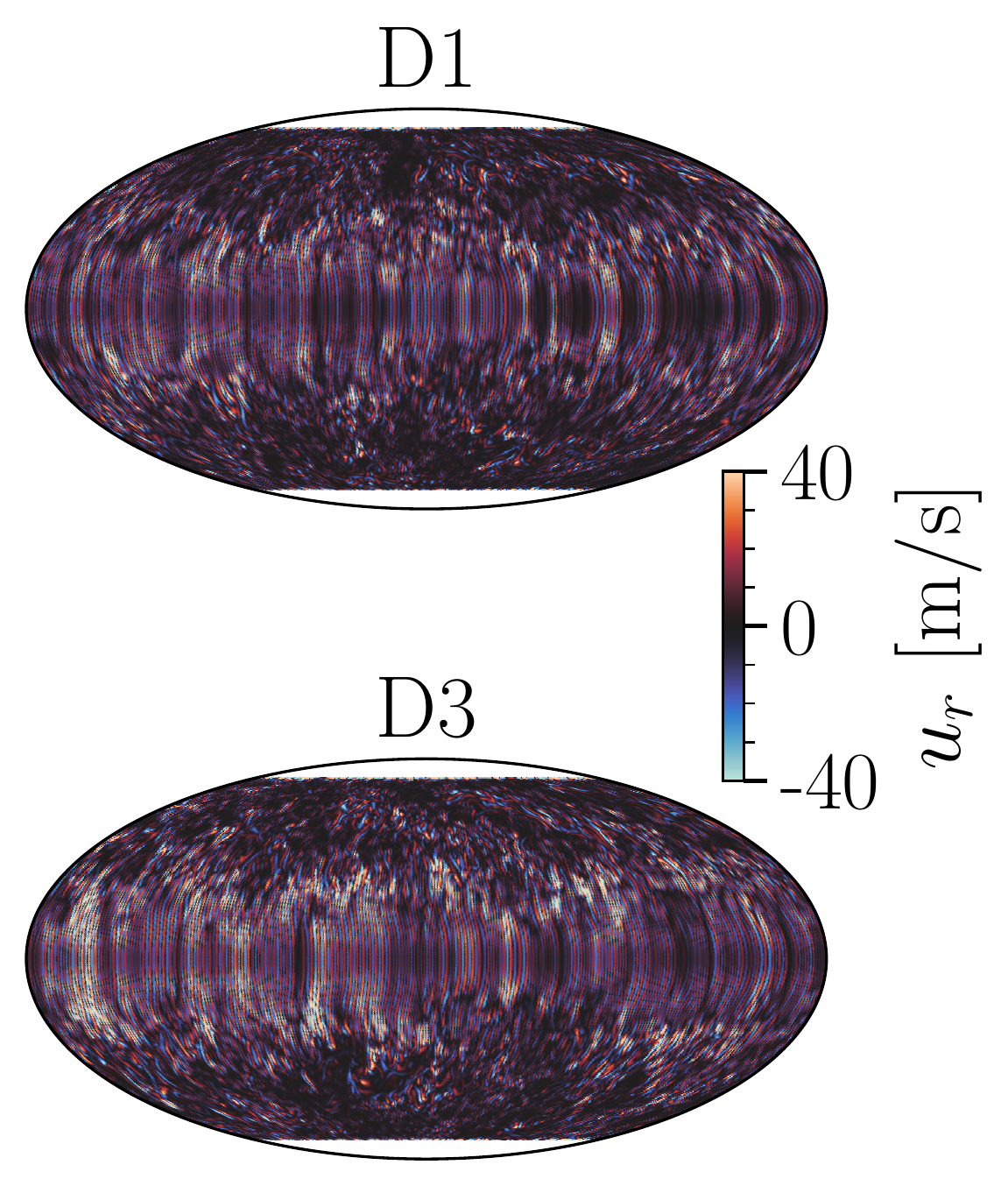}
        \caption{Mollweide projections of radial velocity near the surface for runs
                D1 and D3.}
        \label{fig:urSurface}%
\end{figure}

\begin{figure}
        \centering
        \includegraphics[width=\linewidth]{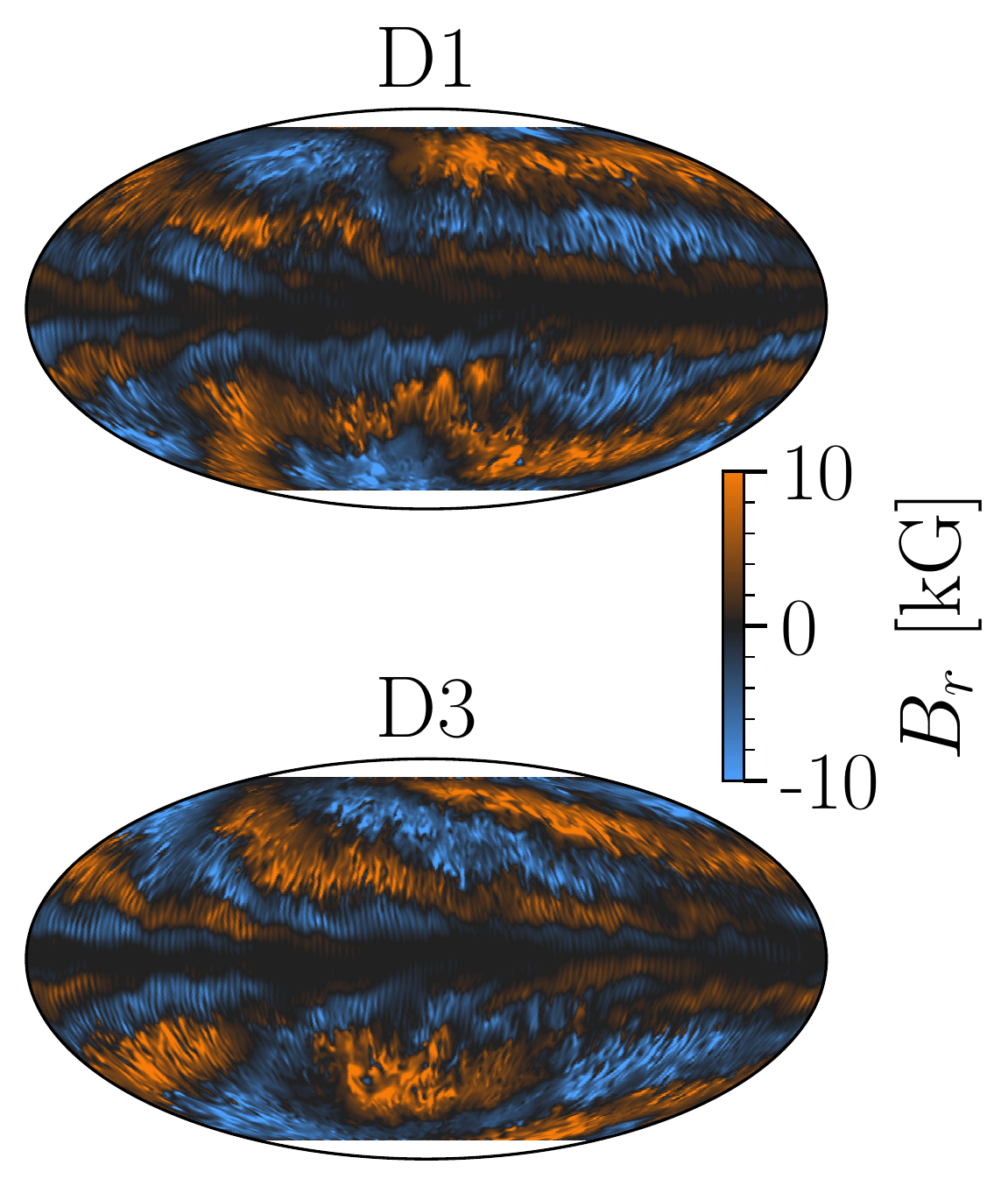}
        \caption{Mollweide projections of radial magnetic field near the surface for runs
                D1 and D3.}
        \label{fig:BrSurface}%
\end{figure}

We present the results of two sets of three simulations each. Set C is
characterized by a rotation rate of $20\Omega_\odot$ and set D by
$30\Omega_\odot$.
Runs C1 and D1 correspond to the parent runs without centrifugal force from
which C2 and C3, and D2 and D3 were forked, respectively. The last four runs
were initialized
with the centrifugal force. For runs C2 and C3, we considered $\mathcal{F}=0.875,$
but they were initialized from C1 at different times. This was done to test whether
the initial magnetic state of the parent run alters the solution of the forked
run. Runs D2 and D3 have $\mathcal{F}=0.875$ and $8.75$, respectively.
This last run is
considered as an extreme case where we exaggerated the effect of the centrifugal
force to show the corresponding implications.
Each simulation had a resolution of $144\times288\times576$
grid points
in $(r,\theta,\phi)$.

\subsection{Dynamical state in the simulations}

In our simulations, the azimuthally averaged density profile
at the equatorial plane of the star 
is basically unaffected by the centrifugal force; the only change we see 
occurs at high latitudes. We focus here on  relative density differences between the 
region $60^\circ$ above the equator and the density profile at the equator, 
which we define via
        \begin{equation}
        \overline{\rho}_{\rm diff} = \frac{\overline{\rho}(90^\circ-\theta=60^\circ) -
        \overline{\rho}(90^\circ-\theta=0^\circ)}{\langle\overline{\rho}\rangle_{r\theta}},
        \end{equation}
where $\langle\overline{\rho}\rangle_{r\theta}$ is the volume-averaged
density.
In the context of the enhanced luminosity method, we recall that 
density differences scale as \citep{Brandenburg2005, Kapyla13,Navarrete20}
\begin{equation}
\overline{\rho}_{\rm diff}\propto \mathcal{L}_r^{2/3}.
\end{equation}
This implies that we should multiply with a factor
$\mathcal{L}_r^{-2/3}$ to obtain the expected density difference in a physical 
star. We calculated these density differences and averaged them over the last 80
years of the simulation. They are shown in Fig.~\ref{fig:rhodiff} as a function
of radius. These differences are more relevant in the interior of 
the star, at $70-80\%$ of the stellar radius, where they have a typical 
magnitude of about $10^{-4}$. These density variations show 
clear trends with the centrifugal force, which tends to increase the density 
difference between higher latitudes and equator toward positive values. 
There are marked differences between runs C1 and C2, but not between runs
D1 and D2. A possible explanation for this might be the different
dynamo solutions. In general, runs in set C show dynamos that tend to alternate
between the two hemispheres. This produces an asymmetry on the density field
with respect to the equator. In this case, $\overline{\rho}_{\rm diff}$ increases when
the magnetic field is more concentrated in one hemisphere. This is the case
for runs C1 and C2. The difference comes from the location of the magnetic
field structure, which reduces the local density. This is not the case for runs D1 and D2, however, and so the density profiles are the
same. A similar explanation can be given for run D3.

Snapshots of the final state of the radial (convective) velocity near the surface
of the star are given in Fig.~\ref{fig:urSurface} for simulations D1 and D3,
which are also representative of the other runs within our set of simulations.
The series of runs C and D correspond to fast rotators, and the 
convective cells are therefore very small toward medium to high latitudes, whereas they become elongated near 
the equator. This is a common phenomenon obtained in 
simulations of stellar convection \citep[see, e.g.,][]{Viviani18} and is 
consistent with the Taylor-Proudman balance. The result for D3 is very similar
as for D1, but it is not identical. While both runs were evolved 
until the same time, an identical result is not expected because the dynamics are nonlinear and because the centrifugal force causes 
perturbations within the star. On the other hand, and even though in principle 
the strength of the centrifugal force is quite significant in run D3,
the impact on the flow pattern appears to be relatively minor.

Similar projections, now for the radial component of the magnetic field, are 
presented in Fig.~\ref{fig:BrSurface}. In run D1, clear nonaxisymmetric 
structures are present that extend throughout each hemisphere. 
Nonaxisymmetric structures seem to be somewhat smaller in run D3,
that is, an $m=2$ mode is also present. The amplitude of $B_r$ remains very 
similar.

\begin{figure}
        \centering
        \includegraphics[width=\columnwidth]{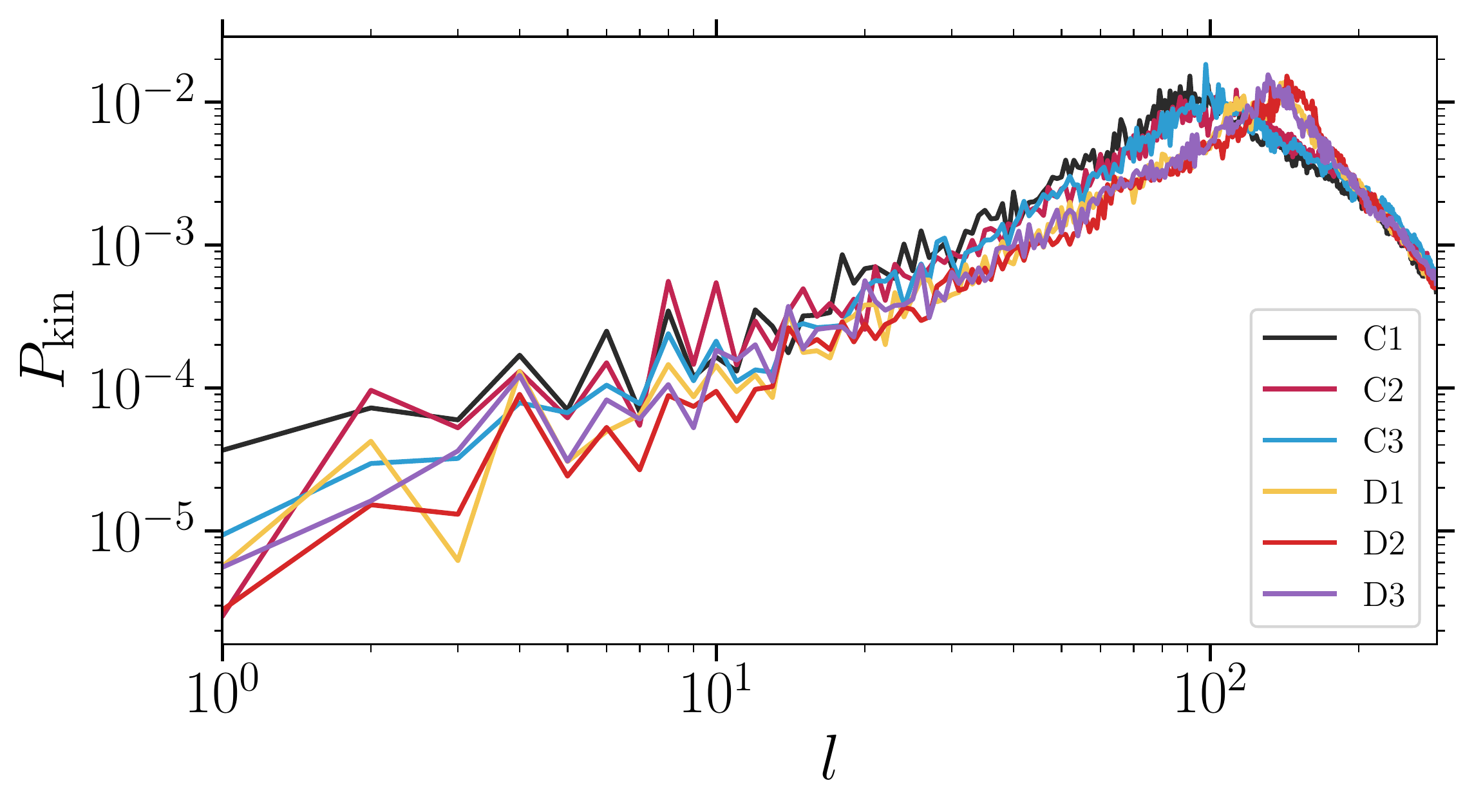}
        \caption{Normalized convective power spectra for runs C1, C2, C3, D1, D2, and 
    D3.}
        \label{fig:ConvPower}%
\end{figure}

We decomposed the radial velocity field at the surface of selected runs
into spherical harmonics and
calculated the normalized convective power spectra as
\begin{equation}\label{eq:convpower}
        P_{\rm kin} = \frac{E_{{\rm kin},\,l}}{\sum_l E_{{\rm kin},\,l}},
\end{equation}
where $E_{{\rm kin},\, l}$ is the kinetic energy of the $l$th degree. This is
shown in Fig. \ref{fig:ConvPower}, where we plot $P$ as a function of $l$ up
to $l_{\rm max} = 288$. This is the maximum resolution we can achieve because we
used 576 grid points along the $\phi$ direction. The convective
power peak is shifted toward higher $l$ for higher rotation rates, but the
centrifugal force has no significant
influence on it, even in the extreme case of run D3. We note that the
contribution of the polar caps, which are not part of the computational domain,
are not included in the spherical harmonic decomposition or in the
power spectra.

\begin{figure}
\centering
\includegraphics[width=\columnwidth]{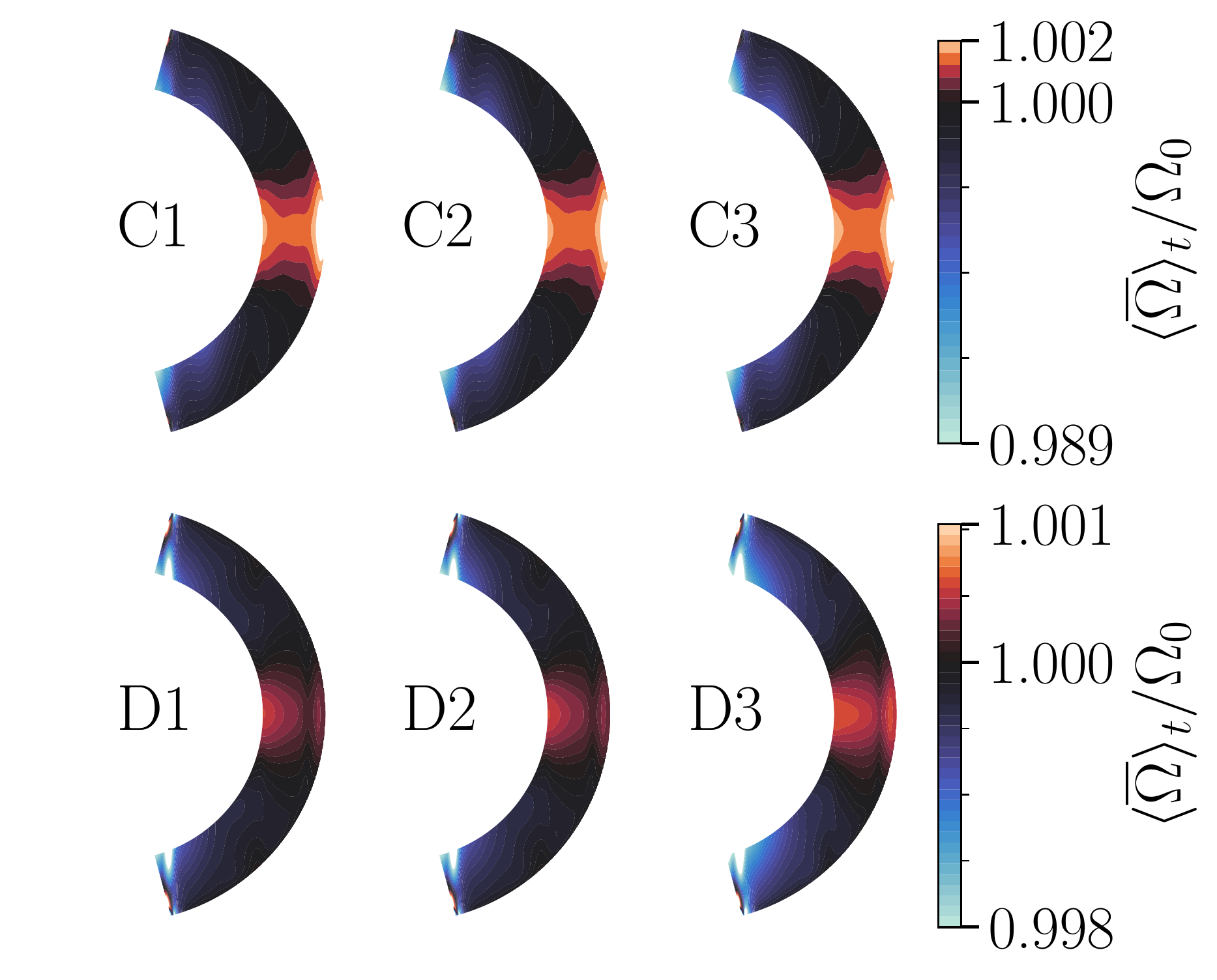}
\caption{Mean rotation rate averaged over the last 80 years of
the simulation and normalized to the rotation rate of the star.
$\langle$...$\rangle_t$ denotes the average over time.}
\label{fig:Omega_C_D}%
\end{figure}

In Fig.~\ref{fig:Omega_C_D} we show the time-averaged and azimuthally averaged angular 
velocity normalized with the angular velocity of the rotating frame for
simulations with different rotation rates and with and without the centrifugal 
force. In set D, where the angular velocity is higher in general, we find that stellar differential rotation is reduced compared with 
set C, which is expected for more rapidly rotating 
runs \citep[e.g.,][]{KR95,Viviani18}. Measures of the radial and latitudinal
differential rotation are shown in Table \ref{tab:runs}. They are defined as
\begin{align}
       & \Delta_\Omega^{(r)} = \frac{\Omega_{\rm eq} -
       \Omega_{\rm bot}}{\Omega_{\rm eq}}, \label{eq:nablar} \\
       & \Delta_\Omega^{(60^\circ)} = \frac{\Omega_{\rm eq} -
       \Omega_{60^\circ}}{\Omega_{\rm eq}}, \label{eq:nablalat}
\end{align}
respectively. Here, $\Omega_{\rm eq}$, $\Omega_{\rm bot}$, and
  $\Omega_{60^\circ}$ are the angular velocities at the equator near
  the surface, at the equator near the bottom of the convective zone,
  and 
near the surface at a latitude of $60^\circ$, respectively. The time-averaged
radial differential rotation
remains practically the same within each set. The time-averaged radial and
latitudinal differential rotation only changes appreciably
in run D3 by a factor of four and two, respectively, where we enhanced the centrifugal force.
We fail to find strong evidence for a direct effect of the 
centrifugal force from the time averages, however.
We also show the rms values of the fluctuations (instantaneous minus average)
of $\Delta_\Omega^{(r)}$ and
$\Delta_\Omega^{(60^\circ)}$ in the last two columns of
Table~\ref{tab:runs}. In general, there is a tendency of increased fluctuations
when the centrifugal force is included and when its amplitude is larger.

Observations of the rapidly rotating K2~dwarf V471~Tau, which is a
PCEB rotating at about 50 times faster than the Sun, show that it
has a solar-like differential rotation \citep{Zaire2022}. The surface
differential rotation is about $\Delta_\Omega^{(60^\circ)}=3.7\times10^{-3}$,
as measured from the shearing of brightness inhomogeneities (Stokes I), and
$\Delta_\Omega^{(60^\circ)}=2.6\times10^{-3}$ from magnetic structures
(Stokes V). The sign of the differential rotation agrees with our
simulations, and the amplitude here is about ten times smaller. We note, however,
that in some cases, the instantaneous value of $\Delta_\Omega^{(60^\circ)}$ can
be as high as $10^{-3}$.

\begin{figure}
\includegraphics[width=\columnwidth]{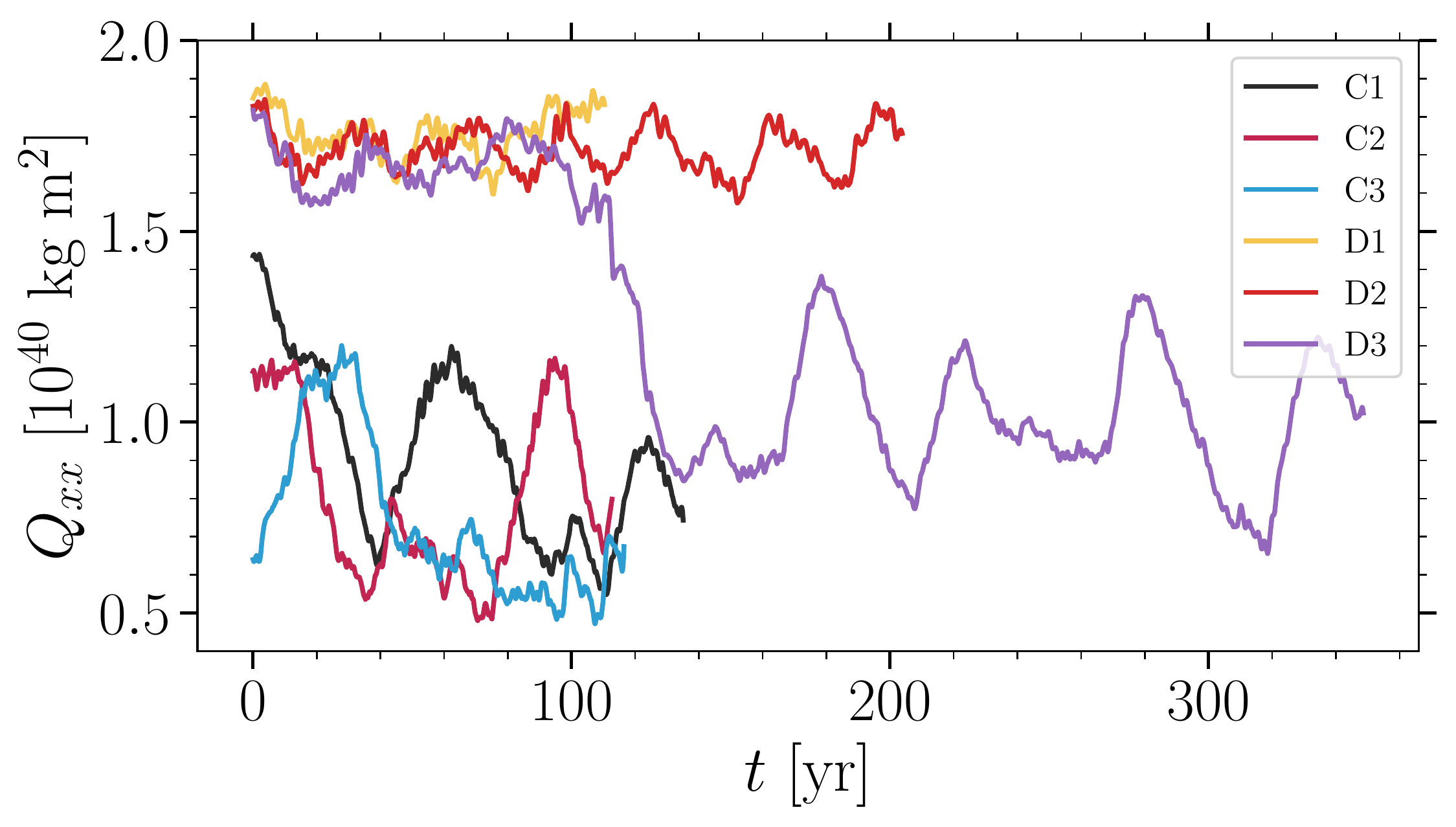}
\caption{$Q_{xx}$ component of the gravitational quadrupole moment as a 
function of time for all runs.}
\label{fig:Qxx}
\end{figure}

\subsection{Gravitational quadrupole moment}

We analyzed the $xx$-component of the gravitational quadrupole moment for 
runs C1, C2, and C3 and for runs D1, D2, and D3. It is defined as
\begin{equation}
Q_{ij} = I_{ij} - \frac{1}{3}\delta_{ij}{\rm Tr} I,
\end{equation}
where
\begin{equation}
I_{ij} = \int\rho({\bm x})x_ix_j{\rm d}V
\end{equation}
is the inertia tensor, with $\rho$ being the density, and $x_i$, $x_j$ are
Cartesian coordinates.
The time evolution of $Q_{xx}$ for all runs is shown
in Fig.~\ref{fig:Qxx}. Their $yy$- and $zz$-components evolve very similarly, 
as demonstrated in \citet{Navarrete20}. The average 
quadrupole moment and its standard deviation are summarized in
Table~\ref{tab:stats} for each simulation. 
\begin{table}
\caption{Summary of the rms value and standard deviation of the 
  quadrupole moment obtained in the simulations. Parentheses indicate powers 
of ten.}
\label{tab:stats}
\centering
\begin{tabular}{c c c}
  \hline\hline
  Run & rms [kg m$^2$] & $\sigma$ [kg m$^2$]\\
  \hline
  C1  & 9.44(39) & 2.19(39) \\
  C2  & 8.25(39) & 2.14(39) \\
  C3  & 7.70(39) & 2.08(39) \\
  D1  & 1.76(40) & 6.38(38) \\
  D2  & 1.71(40) & 5.94(38) \\
  D3  & 1.02(40) & 1.66(39) \\
  \hline
\end{tabular}
\end{table}

\begin{figure*}
\includegraphics[width=\linewidth]{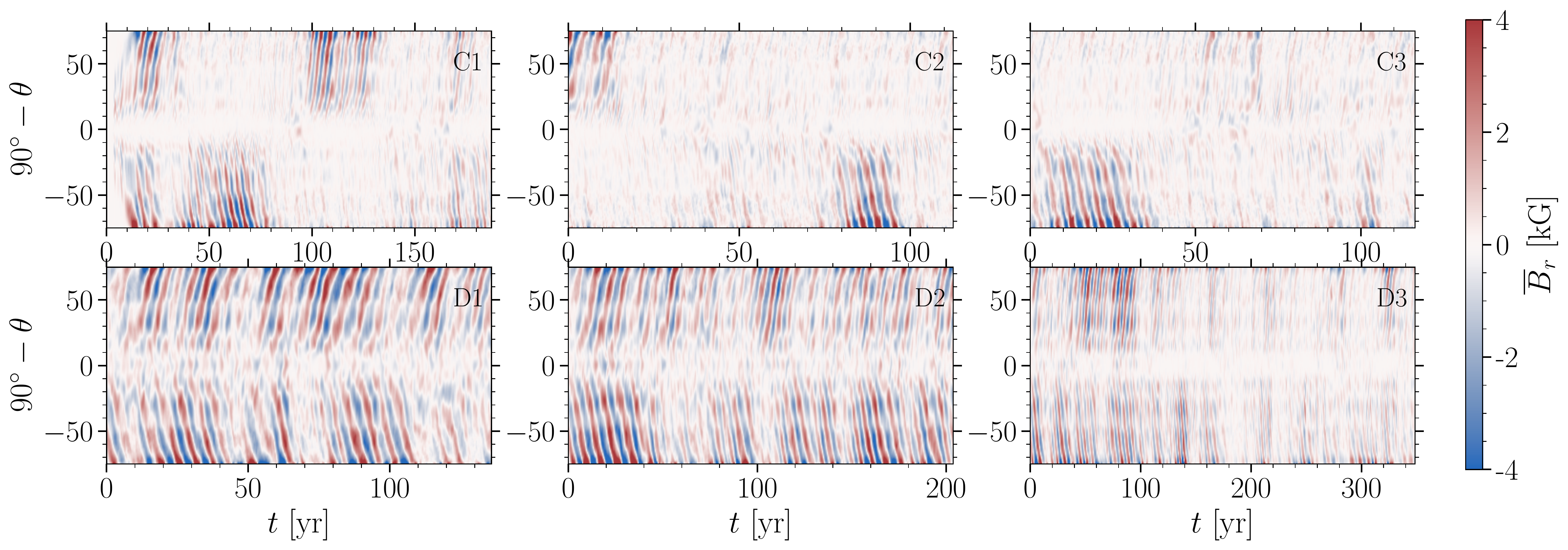}
\caption{Mean (azimuthally averaged) radial magnetic field near the surface for
all runs.}
\label{fig:Br}%
\end{figure*}

For series C with the somewhat lower rotation rate, the average value of the 
quadrupole moment appears to be slightly lower ($\sim7-9\times10^{39}$~kg~m$^2$)
than in series D ($\sim1-2\times10^{40}$~kg~m$^2$), while the standard 
deviation appears to be larger for series C ($\sim2\times10^{39}$~kg~m$^2$) 
than for series D ($\sim6-10\times10^{38}$~kg~m$^2$). Within the range of 
uncertainty, the mean value and the variation appear to be similar within 
the set of simulations C1, C2, and C3, as well as within the set of simulations D1,
D2, and D3. This is to say that the centrifugal force does not appear to affect the mean value of the quadrupole moment very strongly. The 
standard deviation appears to be almost unaffected by the centrifugal 
force within set C. Some more coherent variations are visible in
run D2 and particularly in run D3, where the centrifugal force is the
strongest of all runs. We note further that the drop of the mean
quadrupole moment value in run D3 around $t=100$~yr coincides with a decrease 
in the mean radial magnetic field around the same time, which is shown in
Fig.~\ref{fig:Br}. 
The magnetic field structure and its time evolution is clearly different in 
all simulations, and we also find differences in simulations with and without 
the centrifugal force. It is difficult to assess, however, whether the origin of this 
difference is essentially related to a possible bimodality 
of the solutions or if the centrifugal force specifically introduces a 
different type of behavior.
Overall, the results thus indicate that the stellar quadrupole 
moment is not very sensitive to the centrifugal force.

\section{Discussion and conclusions}

We presented a series of numerical simulations with which we investigated stellar 
dynamos of solar-mass stars with angular velocities of 20 and 30 times the solar 
rotation. The simulations were performed using the enhanced luminosity 
method \citep{Brandenburg2005, Kapyla13, Navarrete20} to avoid
prohibitively large gaps in the relevant timescales. This entails the
use of correspondingly enhanced rotation rates to ensure a realistic
Coriolis number in the simulations, which is required to reach
realistic magnitudes of the drivers of dynamo action, such as the $\alpha$
and $\Omega$ effects. We included and varied
the strength of the centrifugal force in these simulations, including 
cases without the centrifugal force or where the strength of the
centrifugal force was enhanced by an order of magnitude.

The centrifugal force in general causes perturbations during the nonlinear 
evolution, so that the models evolve differently in the 
details, although it is hard to identify clear systematic effects. We note in particular that the averaged radial density profile of the stars remains almost
unchanged, while the density difference between the equator and high latitudes
($60^\circ$) changes by a relative amount of $10^{-4}$. We see some difference
between the distribution of axisymmetric and nonaxisymmetric modes of
the dynamo, and the convective power spectra are affected by the 
strength of the angular velocity, but not so much by the centrifugal force.
Except for the behavior of the density difference (Fig.~\ref{fig:rhodiff}), we
found no clear systematic effects that were due to the centrifugal force.

We similarly find that the mean and 
standard deviation of the quadrupole moment depend more strongly on the angular velocity, while 
the influence of the centrifugal force is weak or almost nonexistent, 
even in the simulation in which the centrifugal force term is enhanced by
an order of magnitude. This is highly relevant because in the original
models proposed by
\citet{applegate92}, the centrifugal force term was supposed to 
give rise to the variation in stellar quadrupole moment, while here we 
find similar variations regardless of the presence of the centrifugal
force. This suggests that the centrifugal force plays only a minor
role in causing this variation, as the overall flow
patterns within the star are driven by more complex dynamics resulting from 
the nonlinear evolution of the system. Adopting the parameters of V471~Tau
\citep{Voelschow16} and inserting the quadrupole variations that we find here
into the framework of \citet{applegate92} \citep[see][]{Navarrete22}, that is,
\begin{equation}
\frac{\Delta P}{P} = -9\frac{\Delta Q_{xx}}{Ma^2},
\end{equation}
where ${\Delta P}/{P}$ is the variation of the period of the binary,
$\Delta Q_{xx}$ is the variation of the quadrupole moment, $M$ is the stellar
mass, and $a$ is the binary separation,
we obtain period variations of the order of $10^{-8}$...$10^{-9}$,
whereas the amplitude of the period variation of close binaries is around 
$10^{-6}$...$10^{-7}$ \citep[see, e.g.,][]{Voelschow18}, and about
$8.5\times10^{-7}$ for V471~Tau \citep{Zaire2022}. While the original model
was useful to motivate the possible origin of the fluctuations via magnetic
activity, it appears to have difficulties overall in explaining the observed
magnitude of the variations, and the centrifugal force is unlikely to be the
main driver of the variations.

As in our previous studies, we chose to apply our results to
V471~Tau alone because of the similarities between the extension of
the convective zones of our model and the real K2~dwarf. We can roughly rescale
the gravitational quadrupole moment variations obtained here to a target star
of mass $\tilde{M}$ and radius $\tilde{R}$ by assuming that the 
quadrupole moment scales with the stellar inertial moment, that is,
\begin{equation}\label{eq:scalingq}
\Delta \tilde{Q}_{xx} = \frac{\tilde{M}\tilde{R}^2}{MR^2}\Delta Q_{xx},
\end{equation}
where $M$ and $R$ are the mass and radius of our simulated star.
Choosing the parameters of NN~Ser of $\tilde{M} = 0.111M_\odot$,
$\tilde{R} = 0.149R_\odot$, and $a=0.934R_\odot$ \citep{Voelschow16}
yields
\begin{equation}
\frac{\Delta P}{P} = 2.34\times10^{-9},
\end{equation}
which is a few hundred times lower than the value estimated from
observations \citep{Voelschow16}. This number should be taken with
extreme caution, however. A potential error source is the missing effect
  of rotation in Eq.~(\ref{eq:scalingq}). The value of
$\Delta \tilde{Q}_{xx}$ is a rather crude estimate and results from
self-consistent simulations that will be presented elsewhere.

As in previous work \citep[e.g.,][]{Navarrete20, Navarrete22}, our simulations 
confirm that MHD simulations of stellar dynamos naturally 
produce an evolution of the stellar quadrupole moment, including one varying and one approximately constant component. The time-varying component may be 
somewhat too small to explain the observed ETVs. 
As suggested in models by \citet{Applegate89} and \citet{Lanza20}, a 
roughly constant component might produce the variations caused by spin-orbit 
coupling via libration or circulation
\citep[see also the discussion in][]{Navarrete22}. As our simulations indeed show a mean and a fluctuating part of the quadrupole moment, the 
observed real variation may well consist of a superposition of the two components,
where the relative strength may depend on the specific system and its 
parameters. As has been demonstrated by \citet{Navarrete20}, the presence of 
magnetic fields is crucial because purely hydrodynamic simulations only produce 
short-term variations on the sound-crossing timescale of the star, but no 
longer-term variations on timescales of years or decades.

From the results obtained here, we thus conclude that neither the
thin-shell model by \citet{applegate92} nor the finite-shell model by
\citet{brinkworth06} correctly describes the origin of the quadrupole moment 
variation because they both assume it to originate in the centrifugal force and 
in an internal redistribution of the angular velocity, while in our simulations, 
a change in centrifugal force term does not lead to any appreciable change in 
the mean or fluctuating component of the quadrupole. The physics causing these 
variations thus requires modeling the stellar dynamo with compressible 
MHD in three dimensions.

Another possibility is given by hybrid solutions
in which ETVs receive contributions of magnetic origin and
from planets. \citet{MaiMutel22} considered such a scenario for three
PCEBs as neither planets nor the Applegate mechanism can fully
account for ETVs. We conclude by encouraging studies of the detection of
circumbinary planets around PCEBs as well as Zeeman-Doppler imaging
of their main-sequence components. The combination of these subjects will help us
constrain the planetary orbits and masses, and to understand to which extent
we can use ETVs to study stellar magnetic fields.

\begin{acknowledgements}
We thank the referee for a detailed review and helpful comments that
improved the manuscript.
FHN acknowledges financial support from the DAAD (Deutscher Akademischer
Austauschdienst; code 91723643) for his doctoral studies.
DRGS gratefully acknowledges support by the ANID BASAL projects ACE210002 and 
FB210003, as well as via the Millenium Nucleus NCN19-058 (TITANs). DRGS and CAOR 
thank for funding via Fondecyt Regular (project code 1201280).
PJK acknowledges financial support by the Deutsche
Forschungsgemeinschaft Heisenberg programme (grant No.\ KA 4825/4-1).
\end{acknowledgements}

\bibliographystyle{aa} 
\bibliography{bibliography} 

\begin{thebibliography}{34}
\expandafter\ifx\csname natexlab\endcsname\relax\def\natexlab#1{#1}\fi

\bibitem[{{Applegate}(1989)}]{Applegate89}
{Applegate}, J.~H. 1989, \apj, 337, 865

\bibitem[{{Applegate}(1992)}]{applegate92}
{Applegate}, J.~H. 1992, \apj, 385, 621

\bibitem[{{Applegate} \& {Patterson}(1987)}]{applegate87}
{Applegate}, J.~H. \& {Patterson}, J. 1987, \apjl, 322, L99

\bibitem[{{Beuermann} {et~al.}(2012){Beuermann}, {Breitenstein}, {Debski},
  {Diese}, {Dubovsky}, {Dreizler}, {Hessman}, {Hornoch}, {Husser}, {Pojmanski},
  {Wolf}, {Wo{\'z}niak}, {Zasche}, {Denk}, {Langer}, {Wagner}, {Wahrenberg},
  {Bollmann}, {Habermann}, {Haustovich}, {Lauser}, {Liebing}, \&
  {Niederstadt}}]{beuermann12}
{Beuermann}, K., {Breitenstein}, P., {Debski}, B., {et~al.} 2012, \aap, 540, A8

\bibitem[{{Beuermann} {et~al.}(2013){Beuermann}, {Dreizler}, \&
  {Hessman}}]{beuermann13a}
{Beuermann}, K., {Dreizler}, S., \& {Hessman}, F.~V. 2013, \aap, 555, A133

\bibitem[{{Beuermann} {et~al.}(2010){Beuermann}, {Hessman}, {Dreizler},
  {Marsh}, {Parsons}, {Winget}, {Miller}, {Schreiber}, {Kley}, {Dhillon},
  {Littlefair}, {Copperwheat}, \& {Hermes}}]{beuermann10}
{Beuermann}, K., {Hessman}, F.~V., {Dreizler}, S., {et~al.} 2010, \aap, 521,
  L60

\bibitem[{{Bours} {et~al.}(2016){Bours}, {Marsh}, {Parsons}, {Dhillon},
  {Ashley}, {Bento}, {Breedt}, {Butterley}, {Caceres}, {Chote}, {Copperwheat},
  {Hardy}, {Hermes}, {Irawati}, {Kerry}, {Kilkenny}, {Littlefair},
  {McAllister}, {Rattanasoon}, {Sahman}, {Vu{\v c}kovi{\'c}}, \&
  {Wilson}}]{bours16}
{Bours}, M.~C.~P., {Marsh}, T.~R., {Parsons}, S.~G., {et~al.} 2016, \mnras,
  460, 3873

\bibitem[{{Brandenburg} {et~al.}(2005){Brandenburg}, {Chan}, {Nordlund}, \&
  {Stein}}]{Brandenburg2005}
{Brandenburg}, A., {Chan}, K.~L., {Nordlund}, {\r{A}}., \& {Stein}, R.~F. 2005,
  Astronomische Nachrichten, 326, 681

\bibitem[{{Brinkworth} {et~al.}(2006){Brinkworth}, {Marsh}, {Dhillon}, \&
  {Knigge}}]{brinkworth06}
{Brinkworth}, C.~S., {Marsh}, T.~R., {Dhillon}, V.~S., \& {Knigge}, C. 2006,
  \mnras, 365, 287

\bibitem[{{Decampli} \& {Baliunas}(1979)}]{decampli79}
{Decampli}, W.~M. \& {Baliunas}, S.~L. 1979, \apj, 230, 815

\bibitem[{{Hardy} {et~al.}(2015){Hardy}, {Schreiber}, {Parsons}, {Caceres},
  {Retamales}, {Wahhaj}, {Mawet}, {Canovas}, {Cieza}, {Marsh}, {Bours},
  {Dhillon}, \& {Bayo}}]{Hardy15}
{Hardy}, A., {Schreiber}, M.~R., {Parsons}, S.~G., {et~al.} 2015, \apjl, 800,
  L24

\bibitem[{{K{\"a}pyl{\"a}} {et~al.}(2020){K{\"a}pyl{\"a}}, {Gent}, {Olspert},
  {K{\"a}pyl{\"a}}, \& {Brandenburg}}]{Kapyla20b}
{K{\"a}pyl{\"a}}, P.~J., {Gent}, F.~A., {Olspert}, N., {K{\"a}pyl{\"a}}, M.~J.,
  \& {Brandenburg}, A. 2020, Geophysical and Astrophysical Fluid Dynamics, 114,
  8

\bibitem[{{K{\"a}pyl{\"a}} {et~al.}(2013){K{\"a}pyl{\"a}}, {Mantere}, {Cole},
  {Warnecke}, \& {Brandenburg}}]{Kapyla13}
{K{\"a}pyl{\"a}}, P.~J., {Mantere}, M.~J., {Cole}, E., {Warnecke}, J., \&
  {Brandenburg}, A. 2013, \apj, 778, 41

\bibitem[{{Kitchatinov} \& {R\"udiger}(1995)}]{KR95}
{Kitchatinov}, L.~L. \& {R\"udiger}, G. 1995, \aap, 299, 446

\bibitem[{{Lanza}(2006)}]{Lanza06}
{Lanza}, A.~F. 2006, \mnras, 369, 1773

\bibitem[{{Lanza}(2020)}]{Lanza20}
{Lanza}, A.~F. 2020, \mnras, 491, 1820

\bibitem[{{Mai} \& {Mutel}(2022)}]{MaiMutel22}
{Mai}, X. \& {Mutel}, R.~L. 2022, \mnras, 513, 2478

\bibitem[{{Marsh} \& {Pringle}(1990)}]{marsh90}
{Marsh}, T.~R. \& {Pringle}, J.~E. 1990, \apj, 365, 677

\bibitem[{{Matese} \& {Whitmire}(1983)}]{matese83}
{Matese}, J.~J. \& {Whitmire}, D.~P. 1983, \aap, 117, L7

\bibitem[{{Mustill} {et~al.}(2013){Mustill}, {Marshall}, {Villaver}, {Veras},
  {Davis}, {Horner}, \& {Wittenmyer}}]{Mustill13}
{Mustill}, A.~J., {Marshall}, J.~P., {Villaver}, E., {et~al.} 2013, \mnras,
  436, 2515

\bibitem[{{Navarrete} {et~al.}(2022){Navarrete}, {K{\"a}pyl{\"a}},
  {Schleicher}, {Ortiz}, \& {Banerjee}}]{Navarrete22}
{Navarrete}, F.~H., {K{\"a}pyl{\"a}}, P.~J., {Schleicher}, D. R.~G., {Ortiz},
  C.~A., \& {Banerjee}, R. 2022, \aap, 663, A90

\bibitem[{{Navarrete} {et~al.}(2020){Navarrete}, {Schleicher},
  {K{\"a}pyl{\"a}}, {Schober}, {V{\"o}lschow}, \& {Mennickent}}]{Navarrete20}
{Navarrete}, F.~H., {Schleicher}, D. R.~G., {K{\"a}pyl{\"a}}, P.~J., {et~al.}
  2020, \mnras, 491, 1043

\bibitem[{{Navarrete} {et~al.}(2018){Navarrete}, {Schleicher}, {Zamponi
  Fuentealba}, \& {V{\"o}lschow}}]{Navarrete18}
{Navarrete}, F.~H., {Schleicher}, D.~R.~G., {Zamponi Fuentealba}, J., \&
  {V{\"o}lschow}, M. 2018, \aap, 615, A81

\bibitem[{{Ogilvie} \& {Lin}(2007)}]{OgilvieLin07}
{Ogilvie}, G.~I. \& {Lin}, D.~N.~C. 2007, \apj, 661, 1180

\bibitem[{{Pencil Code Collaboration} {et~al.}(2021){Pencil Code
  Collaboration}, {Brandenburg}, {Johansen}, {Bourdin}, {Dobler}, {Lyra},
  {Rheinhardt}, {Bingert}, {Haugen}, {Mee}, {Gent}, {Babkovskaia}, {Yang},
  {Heinemann}, {Dintrans}, {Mitra}, {Candelaresi}, {Warnecke},
  {K{\"a}pyl{\"a}}, {Schreiber}, {Chatterjee}, {K{\"a}pyl{\"a}}, {Li},
  {Kr{\"u}ger}, {Aarnes}, {Sarson}, {Oishi}, {Schober}, {Plasson}, {Sandin},
  {Karchniwy}, {Rodrigues}, {Hubbard}, {Guerrero}, {Snodin}, {Losada},
  {Pekkil{\"a}}, \& {Qian}}]{PencilCode}
{Pencil Code Collaboration}, {Brandenburg}, A., {Johansen}, A., {et~al.} 2021,
  The Journal of Open Source Software, 6, 2807

\bibitem[{{Schleicher} \& {Dreizler}(2014)}]{schleicher14}
{Schleicher}, D.~R.~G. \& {Dreizler}, S. 2014, \aap, 563, A61

\bibitem[{{Vaccaro} {et~al.}(2015){Vaccaro}, {Wilson}, {Van Hamme}, \&
  {Terrell}}]{Vaccaro15}
{Vaccaro}, T.~R., {Wilson}, R.~E., {Van Hamme}, W., \& {Terrell}, D. 2015,
  \apj, 810, 157

\bibitem[{{Viviani} {et~al.}(2018){Viviani}, {Warnecke}, {K{\"a}pyl{\"a}},
  {K{\"a}pyl{\"a}}, {Olspert}, {Cole-Kodikara}, {Lehtinen}, \&
  {Brandenburg}}]{Viviani18}
{Viviani}, M., {Warnecke}, J., {K{\"a}pyl{\"a}}, M.~J., {et~al.} 2018, \aap,
  616, A160

\bibitem[{{V{\"o}lschow} {et~al.}(2014){V{\"o}lschow}, {Banerjee}, \&
  {Hessman}}]{volschow14}
{V{\"o}lschow}, M., {Banerjee}, R., \& {Hessman}, F.~V. 2014, \aap, 562, A19

\bibitem[{{V{\"o}lschow} {et~al.}(2018){V{\"o}lschow}, {Schleicher},
  {Banerjee}, \& {Schmitt}}]{Voelschow18}
{V{\"o}lschow}, M., {Schleicher}, D.~R.~G., {Banerjee}, R., \& {Schmitt},
  J.~H.~M.~M. 2018, \aap, 620, A42

\bibitem[{{V{\"o}lschow} {et~al.}(2016){V{\"o}lschow}, {Schleicher},
  {Perdelwitz}, \& {Banerjee}}]{Voelschow16}
{V{\"o}lschow}, M., {Schleicher}, D.~R.~G., {Perdelwitz}, V., \& {Banerjee}, R.
  2016, \aap, 587, A34

\bibitem[{{Zahn} \& {Bouchet}(1989)}]{zahn89}
{Zahn}, J.~P. \& {Bouchet}, L. 1989, \aap, 223, 112

\bibitem[{{Zaire} {et~al.}(2022){Zaire}, {Donati}, \& {Klein}}]{Zaire2022}
{Zaire}, B., {Donati}, J.~F., \& {Klein}, B. 2022, \mnras, 513, 2893

\bibitem[{{Zorotovic} \& {Schreiber}(2013)}]{zorotovic13}
{Zorotovic}, M. \& {Schreiber}, M.~R. 2013, \aap, 549, A95

\end{thebibliography}

\end{document}